\documentclass[12pt, letterpaper]{JHEP3}
\pdfoutput=1
\usepackage{amssymb, amsmath, amsopn, amsthm}
\usepackage{epsfig}
\usepackage{psfrag}
\usepackage{subfigure}
\usepackage{cite}
\usepackage{epstopdf}
\usepackage{ulem}

\newcommand{\eref}[1]{(\ref{#1})}

\newcommand{\cref}[1]{Chapter~\ref{#1}}
\newcommand{\beq}{\begin{equation}}
\newcommand{\eeq}{\end{equation}}
\newcommand{\ba}{\begin{array}}
\newcommand{\ea}{\end{array}}
\newcommand{\bcenter}{\begin{center}}
\newcommand{\ecenter}{\end{center}}

\def\IB{\relax\hbox{$\inbar\kern-.3em{\rm B}$}}
\def\IC{\relax\hbox{$\inbar\kern-.3em{\rm C}$}}
\def\ID{\relax\hbox{$\inbar\kern-.3em{\rm D}$}}
\def\IE{\relax\hbox{$\inbar\kern-.3em{\rm E}$}}
\def\IF{\relax\hbox{$\inbar\kern-.3em{\rm F}$}}
\def\IG{\relax\hbox{$\inbar\kern-.3em{\rm G}$}}
\def\IGa{\relax\hbox{${\rm I}\kern-.18em\Gamma$}}
\def\IH{\relax{\rm I\kern-.18em H}}
\def\IK{\relax{\rm I\kern-.18em K}}
\def\IL{\relax{\rm I\kern-.18em L}}
\def\IP{\relax{\rm I\kern-.18em P}}
\def\IR{\relax{\rm I\kern-.18em R}}
\def\IZ{\relax\ifmmode\mathchoice
{\hbox{\cmss Z\kern-.4em Z}}{\hbox{\cmss Z\kern-.4em Z}}
{\lower.9pt\hbox{\cmsss Z\kern-.4em Z}}
{\lower1.2pt\hbox{\cmsss Z\kern-.4em Z}}\else{\cmss Z\kern-.4em Z}\fi}
\def\II{\relax{\rm I\kern-.18em I}}

\def\sCC{{\kern 0.27em\vrule height1.45ex width0.03em depth0em
          \kern-0.30em\rm C}}
\def\C{{\mathchoice
  {\sCC}
  {\sCC}
  {\kern 0.225em \vrule height1.05ex width0.025em depth0em \kern-0.25em \rm C}
  {\kern 0.180em \vrule height0.78ex width0.02em depth0em \kern-0.2em \rm C}
        }}
\def\sHH{{\rm I\kern-.16em{}H}}
\def\H{{\mathchoice
  {\sHH}
  {\sHH}
  {\rm I\kern-.13em{}H}
  {\rm I\kern-.13em{}H} }}
\def\sNN{{\rm I\kern-.16em{}N}}
\def\N{{\mathchoice
  {\sNN}
  {\sNN}
  {\rm I\kern-.12em{}N}
  {\rm I\kern-.10em{}N} }}
\def\sPP{{\rm I\kern-.16em{}P}}
\def\P{{\mathchoice
  {\sPP}
  {\sPP}
  {\rm I\kern-.12em{}P}
  {\rm I\kern-.10em{}P} }}
\def\sQQ{{\kern 0.27em \vrule height1.45ex width0.03em depth0em
          \kern-0.30em \rm Q}}
\def\Q{{\mathchoice
        {\sQQ}
        {\sQQ}
  {\kern 0.225em \vrule height1.05ex width0.025em depth0em \kern-0.25em \rm Q}
  {\kern 0.180em \vrule height0.78ex width0.020em depth0em \kern-0.20em \rm Q}
        }}
\def\sRR{{\rm I\kern-0.16em{}R}}
\def\R{{\mathchoice
  {\sRR}
  {\sRR}
  {\rm I\kern-0.12em{}R}
  {\rm I\kern-0.10em{}R} }}
\def\sZZ{{\rm Z\kern-0.32em{}Z}}
\def\Z{{\mathchoice
  {\sZZ}
  {\sZZ} 
  {\rm Z\kern-0.3em{}Z}     
  {\rm Z\kern-0.25em{}Z} }}  
\def\ZZZ{{\rm Z\kern-0.24em{}Z}}
\def\sII{{\rm I\kern-0.16em{}I}}
\def\I{{\mathchoice
  {\sII}
  {\sII}
  {\rm I\kern-0.12em{}I}
  {\rm I\kern-0.10em{}I} }}

\def\Tr{{\rm Tr}}

\def\inbar{\,\vrule height1.5ex width.4pt depth0pt}
\font\cmss=cmss10 \font\cmsss=cmss10 at 7pt

\def\smiley{\hbox{\large$\bigcirc$\hspace{-0.80em}\raise.2ex
\hbox{$\cdot\cdot$}\kern-.61em\lower.2ex\hbox{\scriptsize$\smile$}}\ }
\def\frowny{\hbox{\large$\bigcirc$\hspace{-0.80em}\raise.2ex
\hbox{$\cdot\cdot$}\kern-.635em\lower.2ex\hbox{\scriptsize$\frown$}}\ }

\def\I{{\rlap{1} \hskip 1.6pt \hbox{1}}}

\makeatletter
\let\hangafter\@hangfrom
\makeatother

\newcommand{\drawsquare}[2]{\hbox{%
\rule{#2pt}{#1pt}\hskip-#2pt
\rule{#1pt}{#2pt}\hskip-#1pt
\rule[#1pt]{#1pt}{#2pt}}\rule[#1pt]{#2pt}{#2pt}\hskip-#2pt
\rule{#2pt}{#1pt}}

\newcommand{\beqa}{\begin{eqnarray}}
\newcommand{\eeqa}{\end{eqnarray}}
\newcommand{\be}{\begin{equation}}
\newcommand{\ee}{\end{equation}}
\newcommand{\bea}{\begin{eqnarray}}
\newcommand{\eea}{\end{eqnarray}}
\newcommand{\bA}{\begin{array}}
\newcommand{\eA}{\end{array}}
\newcommand{\bc}{\begin{center}}
\newcommand{\ec}{\end{center}}

\renewcommand{\thepage}{\arabic{page}} 

\newcommand{\Yfund}{\raisebox{-.5pt}{\drawsquare{6.5}{0.4}}}

\def\be{\begin{equation}}
\def\ee{\end{equation}}
\def\bea{\begin{eqnarray}}
\def\eea{\end{eqnarray}}

\def\l{\left}
\def\r{\right}

\title{Metastable Vacua in Warped Throats at Non-Isolated Singularities}

\author{~{\normalsize \bfseries \sffamily Du\v san Simi\' c${}^{1,2}$}

~\\${}^1$Department of Physics, Stanford University\\
Stanford, CA 94305 USA \\ \vspace{0.3cm}

${}^2$ Theory Group, SLAC National Accelerator Laboratory\\
Menlo Park, CA 94025 USA \\ \vspace{0.3cm}

\email{simic@stanford.edu}\\
}

\newcommand{\bbea}{\begin{equation} \begin{aligned}} \newcommand{\eeea}{\end{aligned} \end{equation}}

\abstract{We study the existence of metastable vacua in cascades based on fractional brane configurations at non-isolated singularities preserving ${\cal N}=1$ supersymmetry.  We find that in a large class of models the extra moduli typically generated along such cascades may be stabilized by utilizing special monopole points found recently. We illustrate this in detail for cascades based on the SPP singularity. The supergravity interpretation of these constructions in terms of warped throats with supersymmetry breaking localized near their tips as well as applications to string compactification is discussed. Our constructions are designed to realize a large class of warped throats with supersymmetry breaking localized inside of a highly curved tip region. }

\def\be{\begin{equation}}
\def\ee{\end{equation}}
\def\bea{\begin{eqnarray}}
\def\eea{\end{eqnarray}}

\newcommand{\cG}{\mathcal{G}}

\newcommand{\cN}{\mathcal{N}}

\newcommand{\cW}{\mathcal{W}}

\begin{document}

\section{Introduction}

Warped throats in Type IIB string theory with supersymmetry breaking localized at their tip play an important role in the KKLT construction of de Sitter space, as well as in various top-down constructions of inflation and particle physics \cite{KKLT,KKLMMT, GP,GP2,Gherghetta:2010cj,BDFKSV,MSS}. By gauge-gravity duality such throats can be realized as the renormalization group trajectory of a cascading quiver gauge theory in a supersymmetry breaking vacuum. This allows for the possibility of constructing such throats from considerations in the quantum field theory dual. Considering the calculational ease with which stable/meta-stable supersymmetry breaking vacua have been established in some field theories this may provide a promising route in certain instances.\footnote{Field theory techniques were successfully applied in the construction of globally non-supersymmetric throats in \cite{KST}. They have so far been inadequate in demonstrating the existence of metastability in the Klebanov-Strassler model \cite{KS,DKS}, despite evidence in gravity/string theory\cite{KPV,DKM,McGuirk:2009xx, BGH}.}

A large class of models where such an approach may be useful arises by considering quiver gauge theories obtained from fractional brane configurations at non-isolated singularities. In the small rank, non-cascading regime, numerous examples of supersymmetry breaking have been constructed in this way, reducing the calculation of the stability of the vacuum to a perturbative one in quantum feld theory\cite{AFGM,BMV, ABFK1,ABFK2}. Such models provide "target models" to realize as the end point of a cascading model. The latter is engineered to have a description in terms of supergravity above some scale much larger than the scale of supersymmetry breaking and a calculable field theory description below it.

The question then arises whether the models of \cite{BMV,AFGM, ABFK1,ABFK2} can be ultra-violet completed into cascades while preserving the existence of the metastable state. Because a cascading quiver may be obtained simply as a larger rank version of the target model, and because our best understood example of a cascade, the KS model\cite{KS},  is perfectly self-similar, it is reasonable, at least naively, to believe that the target model may be completed into such a cascade without spoiling the existence of a supersymmetry breaking vacuum. However, recent work implies that the self-similarity assumption of cascades at non-isolated singularities can break down in at least two important ways\cite{Simic1,ABC,BBCC,FHU}.

 From \cite{Simic1,ABC, BBCC, FHU} it is seen in several explicit examples that a cascade based on a non-isolated singularity generically has regions in its renormalization group flow where a gauge coupling associated to a node with an adjoint approaches a strong coupling singularity.\footnote{This is different from what is assumed in \cite{AK}.} The result is a spontaneous breakdown of the gauge group which splits the quiver into two sectors, as was recently shown in \cite{Simic1}. The first sector is identical to the quiver before the transition, except with reduced ranks, while the second consists of a number of $U(1)$ factors and moduli, and in certain cases monopoles. These steps in the cascade therefore break the expected self-similarity by generating additional sectors consisting of Abelian degrees of freedom. Proving that the supersymmetry breaking vacuum of the original target model survives the ultra-violet completion to such cascades thus requires that one demonstrate the additional scalars in these sectors do not lead to runaways.
 
In addition to this potential for dynamical instabilities, there are potential renormalization group instabilities, as noted in \cite{Simic1}. The assumption that the target model always emerges, even if only as a proper sub-sector of a larger effective field theory, may not always be a good one. This can happen if quantum corrections introduce relevant deformations of the renormalization group flow which grow to significance by the time the expected scale of the target model is reached. In such a case the resulting low energy model may differ significantly from the target model, and there is a priori no reason to expect the persistence of a metastable vacuum. 
 
 We will illustrate solutions to these problems in the context of a specific model, that of cascades based on the SPP singularity. A model of supersymmetry breaking based on SPP was recently studied in \cite{BMV}. We expect our analysis of the embedding of this example into a cascade to generalize easily to target models which employ the basic supersymmetry breaking strategy of \cite{BMV}, which is based on a deformation of a given non-isolated singularity with an isolated enhancement, thus giving rise to an infinite class of examples.

The first problem will be solved by making use of the results on adjoint transitions described in \cite{Simic1}. We will see that there is a vacuum in which monopoles condense in such a way that gaps the extra Abelian sectors, thus stabilizing these sectors against a runaway. The second problem will be addressed by showing that it is possible to impose sufficient symmetry so as to forbid the presence of dangerous corrections to the superpotential, while preserving the existence of a supersymmetry breaking vacuum at the end of the cascade.

The organization of this paper is the following. In section 2 we elaborate on the basic setup. In section 3 we discuss the SPP cascade and find that it harbors a metastable vacuum with all moduli stabilized. In section 4 we discuss string compactification. In section 5 we conclude.

\newpage 
\section{Basic Setup}

The construction of warped throats experiencing localized supersymmetry breaking at their tips through their quantum field theory duals sounds contradictory, as it is well known that whenever the gravity description is weakly curved and calculable, the field theory description is incalculable, except for a certain set of protected observables. Since the existence of a metastable vacuum is in general not inferable from such observables (by any known means), it should thus follow that field theory techniques are likely to be of little help in constructing examples.

\begin{figure}[h]
\begin{center}
\epsfig{figure=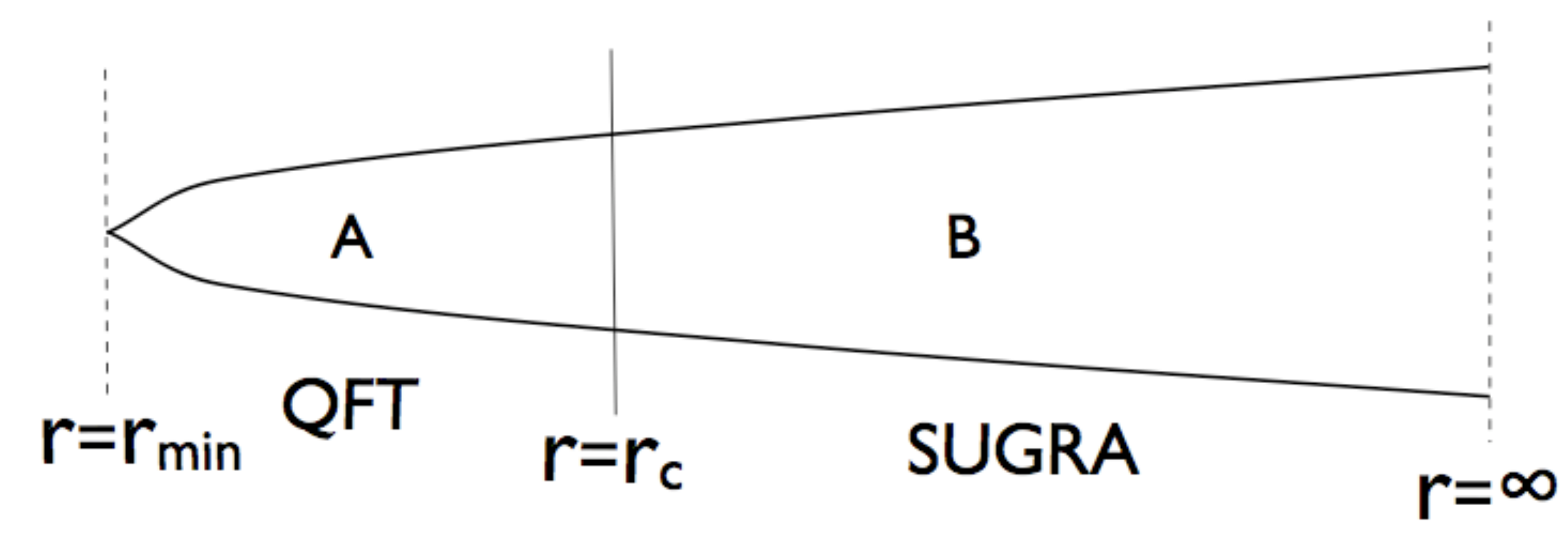,scale=.40,angle=0}
\vspace{-2mm}
\end{center}
\noindent
\caption{\it  In region B supergravity is valid. As region A is approached the geometry becomes more and more curved. Within region A the geometry is highly curved but perturbative quantum field theory applies. The supersymmetry breaking occurs deep within region A where a perturbative QFT calculation may be used to estabilish stability. The model is typically gapped with some mass gap $m_{\rm gap} = r_{\rm min}/\alpha'$.}
\label{throat}
\end{figure}

However, this line of reasoning has an important loop-hole. The loop-hole occurs when two conditions are met: 

\vspace{3mm}

1) The supergravity approximation is valid everywhere {\it except} a region enclosing \indent  the bottom from which the supersymmetry breaking is sourced.
\vspace{3mm}

2) The scale at which the supergravity description breaks down is parametrically \indent larger than the scale of supersymmetry breaking. 
\vspace{3mm}

\noindent The geometry of such a  throat is depicted in figure \ref{throat}. One may then hope to derive the low energy effective field theory governing the highly curved region using the power of supersymmetric field theory techniques alone or in combination with SUGRA and then demonstrate, using that field theory, that it harbors a metastable/stable supersymmetry breaking vacuum. 

In fact, we will find strong evidence that such an approach can be successfully implemented for a large class of examples based on the example of \cite{BMV} and the generalizations advocated there. 

It should be noted that around the region $r_c$, neither field theory nor supergravity is a good description. This loss of control may or may not be dangerous, and is briefly discussed at the end of \S3.4.

\section{Metastable Cascades}

Here we examine the existence of metastable vacua in cascades based on non-isolated singularities. 

\subsection{Quiver}

Consider the theory with gauge group $\cG =  U(N)_1 \times U(N+M_2)_2 \times U(N+M_3)_3 $  and matter content and superpotential:

 \begin{equation}
\label{SPPmatter}
\begin{array}{c|ccc}
    \cG & U(N)_1 & U(N+M_2)_2 & U(N+M_3)_3  \\ \hline
 \Phi & \rm adj   & 1 & 1  \\
  X, \tilde X  &  \Yfund, \overline \Yfund  & 1 & \overline \Yfund, \Yfund  \\
  Y, \tilde Y  & \overline \Yfund, \Yfund  &  \Yfund, \overline \Yfund & 1    \\
  Z, \tilde Z  & 1 & \overline \Yfund,  \Yfund   & \Yfund, \overline  \Yfund    \\
\end{array}
\end{equation}

\be
\cW = \Tr\{ \Phi \tilde Y Y \} - \Tr \{ \Phi X \tilde X\} + \Tr \{ Z \tilde Z \tilde X X - \tilde Z Z Y \tilde Y \}  - \xi \Tr \{ \Phi - \tilde Z Z\}.
\label{SPPsuperpot}
\ee

This model arises as the world volume gauge theory of $N$ D3 branes and some number of fractional-D3 branes at a deformation of the SPP singularity \cite{BMV}. The undeformed version is obtained in the limit $\xi \rightarrow 0$.\footnote{When $\xi = 0$ and all the ranks are equal the theory is superconformal. The resulting super-conformal $U(1)_R$ charge assignments are:
\be
\label{R_charges}
\begin{array}{c|cc}
         & &U(1)_R  \\ \hline
   \Phi &\quad  &2 - 2/\sqrt{3} \\
   X, \tilde X & & 1/\sqrt{3}  \\
   Y , \tilde Y& &1/\sqrt{3}  \\
   Z, \tilde Z & &1-1/\sqrt{3}  \\ 
 \end{array} 
\ee
These may be computed using a-maximization \cite{IW}. }

\subsection{Cascade}

We consider starting the theory off at some ultraviolet scale $\mu_0$ with large and nearly equal ranks. The parameters are chosen such that around $\mu_0$ the theory is well within the supergravity regime.\footnote{These two conditions require $1 >> M_i/N$ and $1 >> g_{\rm YM}^2 >> 1/N$.} 

In order to retain better control over the cascade, we will restrict our starting rank configurations to the form:  $(N,N-qM,N-pM)$ for some co-prime integers $q > p$. For this rank configuration node 1 can be arranged to reach strong coupling first, resulting in an adjoint transition.

\vspace{3mm}
\noindent { \bf Adjoint Transitions }
\vspace{3mm}

We will follow the prescription for adjoint transitions developed in \cite{Simic1,BBCC,ABC}, and implement them by approximating the adjoint node with a copy of $\cN=2$ SQCD, choosing a point on its Coulomb branch. For a more in depth discussion and motivation we refer the reader to \cite{Simic1,BBCC,ABC}. A detailed justification for this prescription in the case of the SPP cascade is beyong the scope of this work.\footnote{  Here we give two simple heuristic motivations: i) from the string theory/geometric description the non-isolated piece of the singularity locally preserves $\cN=2$ thus providing a geometric motivation for the appearance of an effective $\cN=2$ dynamics in an $\cN=1$ set-up\cite{Simic1,BBCC,ABC}; ii) as in the field theory analysis of \cite{Simic1} we expect the supergravity regime to be holomorphically connected to a 'field theory' regime where the effective $\cN=2$ dynamics is manifest. By the standard lore that supersymmetric field theories undergo no phase transitions, elements of the $\cN=2$ description survive upon analytic continuation to the supergravity regime\cite{Intriligator:1994sm}.}

The vacuum structure of $\cN=2$ SQCD is analyzed in detail in \cite{APS}. A point in its space of vacua which will play a special role in our analysis is the baryonic root. For clarity, we briefly review the properties of $\cN=2$ SQCD at this point. Assume the UV theory has $N_c$ colors and $N_f$ quarks. Written in $\cN=1$ components, the UV superpotential is:

\be
\cW_{\rm UV} = \sqrt{2} \sum_{i=1}^{N_f} \Tr \{ \Phi Q^i \tilde Q_i \}
\ee
where $\Phi$ is the adjoint chiral superfield partnered with the gluons of $U(N_c)$ through $\cN=2$ supersymmetry, and $\{ Q^i, \tilde Q_i\}_{i=1}^{N_f}$ comprise $N_f$ quark hypermultiplets. The infrared theory at the baryonic root has $\tilde N_c = N_f-N_c$ colors and $N_f$ quarks. In addition there is a $U(1)^{2N_c-N_f}$ worth of photons coupled to $2N_c-N_f$ monopole hypermultiplets. Written in $\cN=1$ components, the IR superpotential is:

\be
\cW_{\rm IR}/\sqrt{2} = \sum_{i=1}^{N_f} \Tr \{ \phi q^i \tilde q_i  \}+ \frac{1}{\tilde N_c} \Tr \{ q^i \tilde q_i \} \sum_{k=1}^{2N_c-N_f} \psi_k - \sum_{k=1}^{2N_c-N_f} \psi_k e_k \tilde e_k  + \frac{1}{\tilde N_c} \Tr \{ \phi \} \sum_{k=1}^{2N_c-N_f} e_k \tilde e_k
\label{broot}
\ee 
where $\phi$ is an adjoint chiral superfield partnered to the gluons of $U(\tilde N_c)$,  the $\psi_k$ are singlets partnered to the photons of $U(1)^{2N_c-N_f}$, the $\{ q^i, \tilde q_i\}$ are the quarks,  and the $\{e_k, \tilde e_k\}$ are the monopoles.

We now return to the discussion of the cascade.  Approximating the adjoint node by $\cN=2$ SQCD and choosing the baryonic root for the adjoint transition, we have a theory with gauge group $U(N')_1\times U(N'+pM)_2\times  U(N'+qM)_3\times U(1)^{pM+qM}$, where $N' = N-pM-qM$, and effective superpotential:
 
\begin{eqnarray}
\cW_{\mu < \mu_1} =&& \Tr\{ \phi \tilde y y \} - \Tr \{ \phi x \tilde x\} + \frac{1}{N'} \l ( \Tr \{\tilde y y\} - \Tr \{x \tilde x\} \r ) \displaystyle{\sum_{k=1}^{pM+qM}} \psi_k   + \frac{1}{N'} \Tr \{ \phi \} \sum_{k=1}^{pM+qM}  e_k\tilde e _k  \nonumber \\ && - \displaystyle{\sum_k} \psi_k e_k \tilde e_k + \Tr \{ Z \tilde Z \tilde x x  - \tilde Z Z y \tilde y \}   - \xi \Tr \{ \phi - \tilde Z Z\}.
\label{adjSup}
\end{eqnarray}
Here $\mu_1$ denotes the scale of the transition, and the rest of the notation is as in \eref{broot}. We perform the field redefinition:

\begin{eqnarray}
\phi \rightarrow && \phi - \frac{1}{N'}\sum_{k=1}^{pM+qM} \psi_k \cdot 1_{N'\times N'} \nonumber \\
 \psi_k \rightarrow&& \psi_k + \frac{1}{pM+qM} \Tr \{ \phi \}
\end{eqnarray}
The new superpotential is:
\begin{eqnarray}
\cW_{\mu < \mu_1} =&& \Tr\{ \phi \tilde y y \} - \Tr \{ \phi x \tilde x\}  + \Tr \{ Z \tilde Z \tilde x x  - \tilde Z Z y \tilde y \}  - \xi \Tr \{ \phi - \tilde Z Z\}    \nonumber \\ && + ( \frac{1}{N'} -\frac{1}{pM+qM}) \Tr \{ \phi \} \sum_{k=1}^{pM+qM}  e_k\tilde e _k   \nonumber \\ &&  - \displaystyle{\sum_{k=1}^{pM+qM}} \psi_k e_k \tilde e_k  -\frac{1}{N'} \displaystyle{\sum_{k=1}^{pM+qM} \psi_k  \sum_{k'=1}^{pM+qM} e_{k'} \tilde e_{k'}}   + \xi \sum_{k=1}^{pM+qM} \psi_k  .
\label{adjSup}
\end{eqnarray}
It is natural to identify the monopoles with wrapped fractional-D3 branes. Their only interaction with the non-Abelian fields from which the bulk geometry emerges is through the term on the second line. In the cascading regime:
\begin{eqnarray}
\Delta_{\phi} &&= 3 - \sqrt{3} + O(M/N') \nonumber \\ && \approx 1.27
\label{deltaphi}
\end{eqnarray}
Thus this interaction is irrelevant.\footnote{To see this note that i) the monopoles are gauge singlets with respect to the strongly interacting non-Abelian sector of the theory and ii) up to corrections of order $M/N$, the theory is at a fixed point. Thus to a good approximation we have the unitarity bound $\Delta_{e_k},\Delta_{\tilde e_k} \geq 1$ which when combined with \eref{deltaphi} implies that the coupling of the monopoles to the non-Abelian sector through \eref{adjSup} is irrelevant. It follows that at energies below $\mu_1$ the behavior of the monopoles should be that of free fields with canonical dimensions.} The irrelevance of this interaction predicts that the wrapped branes have a wave-function $\Psi_k(r)$ which falls off as:

\be
\Psi_k(r) \sim \l(\frac{r}{\mu_1\alpha'} \r)^{.27} , \quad r << \mu_1 \alpha'
\ee
where $r$ is the radial coordinate in what would be the Poincare patch of AdS if the theory were exactly conformal. Happily, this is consistent with the expectation that particle wave-functions are localized in the supergravity regime around the radial position corresponding to their compositeness scale \cite{PP}.

For simplicity we will neglect to include the effect of this coupling as well as the second to last term in \eref{adjSup}. The first is irrelevant throughout the cascade, while the second is $1/N'$ suppressed relative to the other terms involving the moduli. This will not significantly effect our analysis of the infrared model in \S3.4.

\vspace{3mm}
\noindent { \bf Seiberg Transitions}
\vspace{3mm}

The next transition can be expected to happen at either nodes 2 or 3. Assuming that node 3 is first to reach strong coupling, we replace it by it's Seiberg dual\cite{Seiberg}. Integrating out all massive mesons, the resulting non-Abelian sector of the theory has  matter content: 

 \begin{equation}
\label{SPPmatter2}
\begin{array}{c|ccc}
    \cG & U(N''-pM)_1 & U(N'')_2 & U(N'' -qM)_3  \\ \hline
 \Phi &  1   & \rm adj & 1  \\
  z, \tilde z  &  \Yfund, \overline \Yfund  & 1 & \overline \Yfund, \Yfund  \\
  X, \tilde X  & \overline \Yfund, \Yfund  &  \Yfund, \overline \Yfund & 1    \\
  Y, \tilde Y  & 1 & \overline \Yfund,  \Yfund   & \Yfund, \overline  \Yfund    \\
\end{array}
\end{equation}
where $N'' = N'+pM$. In addition there is a $U(1)^{pM+qM}$ factor with $pM+qM$ moduli and monopoles as described above. The effective superpotential at scales below the Seiberg transition, $\mu_2$, is:

\begin{eqnarray}
\cW_{\mu < \mu_2} &=& \Tr\{ \Phi \tilde Y Y \} - \Tr \{ \Phi X \tilde X\} + \Tr \{ z \tilde z \tilde X X - \tilde zz Y \tilde Y \}  + \xi \Tr \{ \Phi - \tilde zz  \}   \nonumber \\ &&  - \displaystyle{\sum_k} \psi_k e_k \tilde e_k   + \xi \sum_{k=1}^{pM+qM} \psi_k.
\end{eqnarray}
The relationship between the fields before and after the transition are $\Phi := \tilde Z Z$ and $X, \tilde X := y, \tilde y$, while $\{Y, \tilde Y\}$, $\{z ,\tilde z \}$ are dual quarks to $\{ Z, \tilde Z \}$, $\{ x,\tilde x\}$ respectively, introduced during the Seiberg duality. The theory is now in an identical form as before the first transition except with $N\rightarrow N''$, a permutation in the labeling of the nodes - $(1,2,3) \rightarrow (3,1,2)$ - and an irrelevantly coupled Abelian sector which we have suppressed. 

The cascade thus continues alternating in a semi-periodic manner between Seiberg dualities and Higgsing. At each adjoint transition we choose the baryonic root.

\subsection{RG stability}

Here we examine the extent to which the description of the cascade in \S3.2 is stable under RG flow. An anomalous $U(1)_R \times U(1)_{R'}$ R-transformation will play an important role in the analysis. The charges of the fields and  $\xi$ are:

 \begin{equation}
\label{symmetry}
\begin{array}{c|c|c}
    \cG & U(1)_R & U(1)_{R'} \\ \hline
 \Phi             & 0   & 2  \\
  X, \tilde X  & 1 & 0  \\
  Y, \tilde Y  & 1  & 0  \\
  Z, \tilde Z  & 0 & 1   \\
  \xi & 2 & 0  \\
\end{array}
\end{equation}
This is a symmetry at the classical level, broken down to a $\mathbb{Z}_M\times \mathbb{Z'}_M$ subgroup at the quantum level. This symmetry group is preserved by cascades whose adjoint transitions pass through the baryonic root\cite{APS}. 

This modest symmetry group is already powerful enough to rule out several dangerous quantum corrections:

\begin{itemize}

\item Consider corrections which generate a potential for the moduli with $\xi=0$:

\be
\displaystyle{\int} d^2 \theta ~\delta W(\psi_k)
\ee
Such corrections are charged under $\mathbb{Z}_M \in \mathbb{Z}_M\times \mathbb{Z}_{M}'$ and hence vanish identically. Thus the moduli produced at each adjoint transition continue to parameterize flat directions. The same reasoning applies to corrections which are also a function of the adjoint field $\phi$. Corrections with $\xi \neq 0$ we expect to be unimportant until scales which are non-perturbatively small in comparison to $\xi^{1/2}$ are reached, by which time the theory in the supersymmetry breaking vacuum will already be gapped.

\item Consider corrections  to the monopole and quark bilinears:

\be
 \displaystyle{\int} d^2 \theta ~ \l(  \displaystyle{\sum_{k=1}^{pM+qM} } m_k e_k \tilde e_k + m_{x \tilde x} \Tr \{ x \tilde x \} + m_{y \tilde y} \Tr \{ y \tilde y\}  \r) ,
\label{qmbilinear}
\ee
where $m_k, m_{x \tilde x},$ and $m_{y \tilde y}$ are a priori general holomorphic functions of gauge invariant combinations of the moduli and adjoint field. Naively $\mathbb{Z}_M'$ invariance implies that:

\begin{eqnarray}
m_k \sim&& \psi \cdot f_k (\psi^M, \phi\psi^{-1}, \phi^{M} ), \nonumber \\
m_{x \tilde x} \sim&& \psi \cdot f_{x \tilde x} (\psi^M, \phi\psi^{-1}, \phi^{M} ), \nonumber \\
m_{y \tilde y} \sim&&  \psi \cdot f_{y \tilde y} (\psi^M, \phi\psi^{-1}, \phi^{M}),
\end{eqnarray}
ie: there is no constant piece in $m_k, m_{x \tilde x}, m_{y, \tilde y}$ and hence no correction to the mass. However, the generator of $\mathbb{Z}_{M}'$ acts through a combination of an R-symmetry and a discrete gauge transformation:

\begin{eqnarray}
\{ \psi_1, \psi_2, \dots,\psi_{aM-1}, \psi_{aM} \} \rightarrow&& e^{4 \pi i /M} \{ \psi_{aM-2a+1} , \dots , \psi_1,\dots, \psi_{aM-2a} \}, \nonumber \\
\{ e_1,e_2, \dots, e_{aM-1}, e_{aM}\} \rightarrow&& \{ e_{aM-2a+1}, \dots, e_1,\dots, e_{aM-2a} \}, \nonumber \\
\{ \tilde e_1,\tilde e_2, \dots, \tilde e_{aM-1}, \tilde e_{aM}\} \rightarrow&& \{ \tilde e_{aM-2a+1},\dots, \tilde e_1, \dots, \tilde e_{aM-2a} \}.
\end{eqnarray}
where $a=p+q$. Quark masses are invariant under the permutation. Thus, while a correction to the quark masses is forbidden, a correction of the form:

\be
\displaystyle{\sum_{k=1}^{pM+qM} } \displaystyle{\int} d^2 \theta ~ e^{ 2 \pi k i/aM } m_0~ e_k \tilde e_k,
\label{monopole}
\ee
where $m_0$ is some c-number, is invariant under $\mathbb{Z}_M'$, and thus allowed  by symmetry. Although at first sight monopole mass corrections appear to make the stabilization mechanism of \S3.4 inviable by gapping the monopoles before they can condense, it is easily seen that there is a simple $\mathbb{Z}_M \times \mathbb{Z}_M'$ symmetric shift in the moduli which eliminates such mass terms. Thus the massless monopole points survive in the moduli space (and in fact for reasons nearly identical to those first discussed in \cite{Simic1}).
\end{itemize}

Thus we have demonstrated that a few of the most dangerous RG instabilities are not present using simple symmetry arguments. It would be interesting to do a more systematic analysis.

\subsection{Infrared model}

We arrange for many cascade steps to occur before scales of order $\xi^{1/2}$ are reached, so that at some scale $\mu_c >> \xi^{1/2}$ the ranks have been depleted to the point where all 't Hooft couplings are small. 

By an appropriate adjustment of the parameters of the cascade it should be possible to arrange for the theory at scales just below $\mu_c$ to have matter content and superpotential:

 \begin{equation}
\label{SPPmatter}
\begin{array}{c|ccc}
    \cG & U(qM-pM)_1 & U(qM)_2 & U(2qM-pM)_3  \\ \hline
 \Phi & \rm adj   & 1 & 1  \\
  X, \tilde X  &  \Yfund, \overline \Yfund  & 1 & \overline \Yfund, \Yfund  \\
  Y, \tilde Y  & \overline \Yfund, \Yfund  &  \Yfund, \overline \Yfund & 1    \\
  Z, \tilde Z  & 1 & \overline \Yfund,  \Yfund   & \Yfund, \overline  \Yfund    \\
\end{array}
\end{equation}
\begin{eqnarray}
\cW_{\mu < \mu_c} &&= \Tr\{ \Phi \tilde Y Y \}  - \Tr \{ \Phi X \tilde X\} + \Tr \{ Z \tilde Z \tilde X X - \tilde Z Z Y \tilde Y \}  - \xi \Tr \{ \Phi - \tilde Z Z\}  \nonumber \\  \nonumber \\ && \quad +  \displaystyle{\sum_{I}\sum_k} \l \{ \xi \psi_k^I  - \psi_k^I e_k^I \tilde e_k^I  \r \}
\end{eqnarray}
with the $I$ summation over adjoint transitions. We further arrange for the strong coupling scales to satisfy $\Lambda_3 >>  \xi^{1/2} >> \Lambda_2$. Following \cite{BMV}, at scales below $\Lambda_3$ we arrive at an infrared-model with non-Abelian matter content and superpotential:

 \begin{equation}
\label{SPPmatter}
\begin{array}{c|ccc}
    \cG & U(qM-pM)_1 & U(qM)_2 & U(0)_3  \\ \hline
 \Phi & 1 & \rm adj  & 1  \\
  Y, \tilde Y  & \overline \Yfund, \Yfund  &  \Yfund, \overline \Yfund & 1    \\
\end{array}
\end{equation}

\begin{eqnarray}
\cW_{\xi^{1/2} < \mu < \Lambda_3} &&= \Tr\{ \Phi \tilde Y Y \}  - \xi \Tr \{ \Phi\} +  \displaystyle{\sum_{I}\sum_k} \l \{ \xi \psi_k^I  - \psi_k^I e_k^I \tilde e_k^I  \r \}
\end{eqnarray}
Due to the condition $\xi^{1/2} >> \Lambda_2$ all 't Hooft couplings are naturally small at scales of order $\xi^{1/2}$. This will allow us to demonstrate the existence of a metastable vacuum in the thus obtained effective field theory. 

At the classical level the effective field theory we have in this way constructed breaks supersymmetry dynamically. The condition:

\be
\frac{\partial \cW }{ \partial \Phi}  =  \tilde Y Y - \xi  = 0,
\ee 
cannot be satisfied due to a mismatch of ranks. The vacuum which minimizes the classical energy is: 

\be
\tilde Y = \l ( \begin{array}{c}   \xi^{1/2}  \cdot 1_{qM-pM\times qM-pM} \\  0_{pM\times qM-pM} \end{array} \r)  , \quad Y^T = \l ( \begin{array}{c}   \xi^{1/2}  \cdot 1_{qM-pM\times qM-pM} \\  0_{pM\times qM-pM} \end{array} \r), \quad \Phi = 0  \nonumber
\label{vac}
\ee

\be
e^I_k =  \xi^{1/2} , \quad  \tilde e_k^I =\xi^{1/2},  \quad \psi_k^I = 0.
\ee
It is easily seen that the condensation of the monopoles gaps the Abelian sector, protecting this sector against quantum induced runaways.

The non-Abelian sector is identical to the model of \cite{BMV}, which in turn is identical to the model of \cite{ISS} except with a gauged flavor group. Since this gauging is weak at scales of order $\xi^{1/2}$ the stabilization of the pseudo-moduli is unchanged from \cite{ISS}, thus resulting in a metastable vacuum. The leading decay channel can be arranged to be as in \cite{ISS} with bounce action\cite{BMV}:\footnote{ To argue this note that there are only two ways in which the leading decay channel could differ from that found in \cite{ISS}. The first is due to physics accessible only through energies or field excursions larger than $\Lambda_3$. Such excursions are larger than those in the process represented by \eref{bounce} and thus should be more suppressed. The second way is due to a small but non-zero $\Lambda_2$, which produces a weak gauging of the "flavor" group. Any additional supersymmetric vacua produced through $\Lambda_2 \neq 0$ must run off to infinite field values in the limit $\Lambda_2 \rightarrow 0$. Thus we expect the bounce action associated with such decays to go like some positive power of $\xi^{1/2}/\Lambda_2  >> 1$, leading to a parametric suppression of such decays. For sufficiently small $\Lambda_2$ these will be subdominant to those represented by \eref{bounce} \cite{BMV}. }

\beq
 S_{\rm bounce} \sim \l (\frac{\Lambda_3}{\xi^{1/2}} \r)^{4(3p-2q)/p} >> 1, 
\label{bounce}
\eeq
leading to an exponentially long-lived vacuum.

Thus we find evidence for the existence of metastable vacua at the bottom of the SPP cascade.

\vspace{3mm}

\noindent {\it Remarks}

\begin{itemize}
\item The assumption that there is a scale $\mu_c$ at which the 't Hooft couplings are perturbative is important for reliably calculating the existence of a long-lived metastable vacuum. Although it can be proven from supergravity that for $g_sM \lesssim 1$ there is a scale at which supergravity no longer applies, this at best guarantees 't Hooft couplings of $O(1)$, which is not the same as the existence of a perturbative regime. This is an important caveat to our derivations. Nonetheless, we find it plausible that as $g_sM$ is tuned towards small values, a perturbative window emerges, and its size increases to infinity as $g_sM \rightarrow 0$, consistent with the expectation that as $g_sM$ is sent to zero, the curvature in an ever increasing region around the tip grows without bound.  
\end{itemize}

\section{String Compactification}

Warped throats are expected to appear quite generically in flux compactifications of string theory down to four space-time dimensions. While the bulk of a general flux compacitifaction is often a dirty, non-analytic solution of the string equations of motion, the warped throat region can often be found analytically, up to corrections whose strength decrease towards the interior. This in principle should allow one to have control over those aspects of the four-dimensional physics which are dominated by the throat.  

Since a given warped throat may appear in a multitude of different flux compactifications, the study of a single throat is in fact the study of a multitude of flux compactifications. This is the gravitational dual of the concept of universality in the quantum field theory. The "gluing in" of a throat into a flux compactification can be regarded as the ultra-violet completion of the quantum field theory dual into a four dimensional theory of quantum gravity. Within a given universality class, the ultra-violet completions differ by irrelevant operators inserted into the Wilsonian effective action at the Planck scale.

A few concrete questions arise with regard to the gluing of the class of throats constructed here into flux compactifications:

\begin{itemize}

\item What is the likely distribution of supersymmetry breaking scales and life-times \indent of such constructions?

\item What is the efficiency with which supersymmetry breaking is transmitted to a \indent probe placed outside of the throat?

\end{itemize}

\noindent A careful examination of these questions in beyond the scope of this work. However, we make a few remarks:

\begin{itemize}

\item In the compactification, the couplings of the field theory dual become dynamical. In principle this allows them to shift so as to restore supersymmetry. However, in a flux compactification with fully stabilized moduli we expect the values of these fields to be frozen, providing a classical barrier to supersymmetry restoration. It would be interesting to understand whether this barrier is larger or smaller than the ones present in the non-compact model.

\item In the flux compactification we expect the value of $\xi$, which sets the value of the supersymmetry breaking scale, to be related to the vacuum expectation of a complex structure modulus\cite{BMV}. Since $\xi$ is fixed by fluxes, and the number of flux configurations is bounded by the topological complexity of CY 4-folds in F-theory - a quantity which itself is thought to be bounded - it follows that there should be non-trivial constraints on $\xi$.  This should then impose a lower bound on the lowest possible (non-trivial) susy-breaking scale, as well as an upper bound on the lifetime, in similar spirit to \cite{FL}.\footnote{We note however that the quantitative upper bound on the lifetime of compactifications based on the throats constructed here may a priori differ substantially from the one derived in \cite{FL}.}

\item Were the throat non-compact, we would know from considerations in the quantum field theory that the strength of supersymmetry breaking in a region located at some $r>r_c$ is bounded above by:

\be
\l( \alpha' \xi^{1/2}/r_c \r)\times (r_c/r)^{4-\Delta}
\label{coupling}
\ee 
where $1<\Delta<4$ is the dimension of the least relevant relevant operator consistent with the symmetries of the supersymmetry breaking vacuum.\footnote{The bound applies to the coefficients of superymmetry breaking operators in the Wilsonian effective action (in which supersymmetry is realized only non-linearly) . The bound is arrived at by estimating the coefficient of the least relevant relevant supersymmetry breaking operator present in the Wilsonian effective action at the RG scale  corresponding to $r/\alpha'$. The first factor arises from the RG flow inside the weakly coupled regime where the anomalous dimensions are small leading to 3 as the dimension of the least relevant relevant operator.} This would shed light on the second question above. However, compactification introduces the effect of the non-normalizable modes becoming normalizable. These modes can in principle transmit the supersymmetry breaking out of the throat, potentially modifying the bound \eref{coupling}.

\end{itemize}

\section{Discussion}

Our main interest in this paper was the construction of warped throats with localized supersymmetry breaking by the identification of cascading quiver gauge theories with supersymmetry breaking vacua. We focused on cascades based on non-isolated singularities, as the smaller rank versions of these theories have been previously identified as harboring supersymmetry breaking vacua\cite{BMV,AFGM,ABFK1,ABFK2}. These provided target models to produce as the endpoint of a cascade.

We identified two potential obstacles to successfully producing the target model or an acceptable variant as the end point of a cascade. The first regarded the existence of additional light fields generated at higher cascade steps. These were understood to potentially lead to runaways in the supersymmetry breaking state. We argued that this problem could be overcome in a large class of models using a monopole-condensation mechanism. The second regarded the generation of relevant operators along the renormalization group flow, providing the possibility of destabilizing the renormalization group trajectory into an unwanted direction. 

We explicitly demonstrated the monopole condensation mechanism, and the ability to impose symmetries so as to prevent some of the most dangerous quantum corrections, in specific examples based on cascades at the SPP singularity. 

A few interesting directions are suggested by this work. The supersymmetry breaking vacua constructed here typically leave some amount of non-Abelian gauge symmetry unbroken along with a diverse spectrum of light fermion species, many charged under the unbroken gauge symmetry. In a theory of particle physics based on string compactifications of our constructions, these would give rise to interesting hidden sectors (which may or may not provide the primary source of supersymmetry breaking). Therefore an interesting direction is to explore the generic features of such sectors and understand the conditions under which they may impact observable phenomena. It would be particularly interesting to make a link with \cite{SZ, BCRWY, EST,JMS}. At a more formal level, it would be interesting to improve the rigor of our analysis in ways suggested in the main text, including questions related to string compactification.\footnote{An interesting application to the physics of the multiverse may be a new test of the bound on the life-time of stringy de Sitter constructions proposed in \cite{BF}, different from the test in \cite{FL}.}

\bigskip
\centerline{\bf{Acknowledgements}}

We wish to thank Rouven Essig, Sebastian Franco, Shamit Kachru and Gonzalo Torroba for discussions. We also thank S. Franco and S. Kachru for a reading of the manuscript and the JHEP referee for their critical feedback. We are supported by the Mayfield Stanford Graduate Fellowship, the Stanford Institute of Theoretical Physics, and by the DOE under contract DE-AC03-76SF00515.

\bigskip

\appendix

\end{document}